\documentclass{article}
\usepackage{spconf,amsmath,tikz,graphicx}
\usepackage{amssymb,amsfonts}
\usepackage{textcomp}
\graphicspath{ {./images/} }

\usepackage[percent]{overpic}

\usepackage{dirtytalk}
\usepackage[labelfont=bf]{caption}

\usepackage[linesnumbered,ruled,vlined]{algorithm2e}
\SetKwComment{Comment}{$\triangleright$\ }{}
\SetKw{KwBy}{by}

\usepackage[flushleft]{threeparttable}
\usepackage{makecell,booktabs}

\usepackage{adjustbox}

\usepackage{tikz}

\usepackage{float}

\title{Unsupervised Region-based Anomaly Detection
in Brain MRI\\ with Adversarial Image Inpainting}

\name{Bao Nguyen$^*$, Adam Feldman$^\ddag$, Sarath Bethapudi$^\dag$, Andrew Jennings$^\dag$, Chris G. Willcocks$^*$}
\address{ $^*$Department of Computer Science, Durham University \\
  $^\ddag$Department of Engineering, University of Exeter \\
  $^\dag$County Durham and Darlington NHS Foundation Trust
}
  
\begin{document}
%
\maketitle
\begin{abstract}
Medical segmentation is performed to determine the bounds of regions of interest (ROI) prior to surgery. By allowing the study of growth, structure, and behaviour of the ROI in the planning phase, critical information can be obtained, increasing the likelihood of a successful operation. Usually, segmentations are performed manually or via machine learning methods trained on manual annotations. In contrast, this paper proposes a fully automatic, unsupervised inpainting-based brain tumour segmentation system for T1-weighted MRI. First, a deep convolutional neural network (DCNN) is trained to reconstruct missing healthy brain regions. Then, upon application, anomalous regions are determined by identifying areas of highest reconstruction loss. Finally, superpixel segmentation is performed to segment those regions. We show the proposed system is able to segment various sized and abstract tumours and achieves a mean and standard deviation Dice score of 0.771 and 0.176, respectively.
\end{abstract}
%
\section{Introduction}
\label{sec:intro}

Brain tumours are a growth of abnormal cells in the brain. Cancerous primary tumours are the most likely to cause severe symptoms, as they spread and cause damage to the surrounding normal brain tissue, raising intracranial pressure. Symptoms range from headaches and nausea to seizures, paralysis, and loss of vision and speech. Additionally, these tumours may cause a hemorrhagic stroke and can result in death within two days. Hence, early discovery is critical to prevent severe degeneration and allows for more treatment options, increasing the chance of recovery.

Segmentation is an important prerequisite to surgery. Manual segmentation requires radiologists to first determine the ROI, then manually to draw the boundaries surrounding the ROI. Although this is possible in 3D, this is usually done in a slice-by-slice manner by encircling the ROI or annotating voxels of interest.  Not only is this time consuming but also prone to subjectivity and results are subject to large intra and inter expert variability, leading to considereable differences in extracted radiomic values \cite{Parmar2014RobustRF}. There is a need for automatic methods to reduce cost, time and bias.

\emph{Contribution.}\hspace{14px}This paper proposes a region-based anomaly detection method using image inpainting to generate a coarse heatmap, followed by a refinement postprocessing stage. Specifically, we adversarially train a DCNN architecture to locate and inpaint missing healthy brain regions. The intuition is that the network will fail to reconstruct unhealthy data that it has not observed in training. We then apply superpixel segmentation and select the segment with the highest reconstruction loss (based on the heatmap), resulting in a automated system that can better capture irregular tumours.

\emph{Related Work.} \hspace{0.2em} Many unsupervised anomaly detection methods operate by training on a set of normal samples only. Anomalies are then determined if they lie outside the manifold of the learned representation. AnoGAN \cite{Schlegl2017UnsupervisedAD} was first to introduce Generative Adversarial Networks (GANs) \cite{Goodfellow2014GenerativeAN} for anomaly detection which was then applied to retinal fluid segmentation in optical coherence tomography images. AnoGAN \cite{Schlegl2017UnsupervisedAD} introduced a new mapping scheme from high-dimensional normal images to the latent space representation (inverse-mapping). Anomalous regions can then be identified by computing the pixel difference between both images which inevitably induces false positives. Manual thresholding is then often required to reduce them. Another drawback of AnoGAN \cite{Schlegl2017UnsupervisedAD} is that the mapping procedure is an expensive iterative process, defined as the minimization through $\gamma=1,2, \ldots, \Gamma$ backpropagation steps, thus requiring $\Gamma$ optmization steps, which result in poor inference-time performance. Following AnoGAN \cite{Schlegl2017UnsupervisedAD}, other works \cite{Zenati2018EfficientGA, Akay2018GANomalySA, Akay2019SkipGANomalySC} have improved upon instance anomaly detection and the computationally expensive mapping procedure. However, these works do not offer the ability to localize anomalous regions.

Inpainting for anomaly detection has been less extensively investigated. Works from \cite{Joung2012ReliableOD, Munawar2015StructuralIO, Haselmann2018AnomalyDU} have been successfully applied to pedestrian segmentation, on-road anomaly segmentation and surface inspection domains. These results indicate success in applications where the domain is consistent and predictable, e.g. with many patterns with small variability. 

\section{Methodology}
\label{sec:methodology}

\begin{figure*}
    \centering
    \fontsize{13}{11}\selectfont
    \scalebox{0.56}{\includegraphics[width=1.2\textwidth]{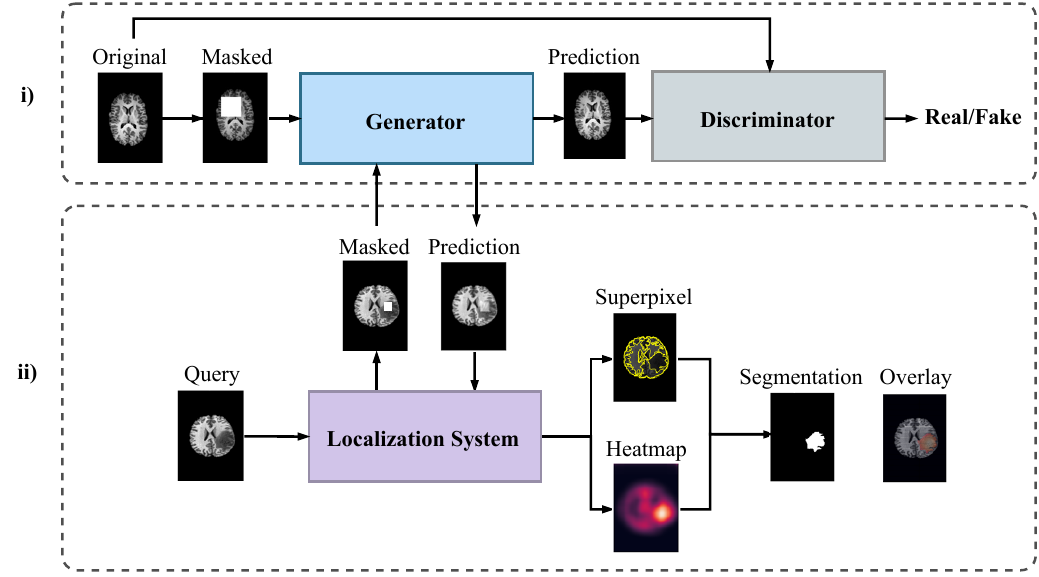}}
    \caption{The Proposed System - i) Train Inpainting Network. ii) Predict anomalous regions and obtain segmentation.}
    \label{fig:System}
\end{figure*}

Brain tumour segmentation is performed in three steps (see Fig. \ref{fig:System}): i) Adversarially train a DCNN to reconstruct missing healthy brain regions. ii) Given a query slice, perform a masked sliding window operation to obtain predictions for all regions. iii) Construct a heatmap indicating areas of highest reconstruction loss and perform superpixel segmentation to better capture object boundary.

The inpainting network consists of two concurrently trained sub-networks, a generator $G$ trained to capture the data distribution of normal brain MRI and a discriminator $D$ trained to classify real slices from $G$'s reconstructions. Both $G$ and $D$ play the minimax two-player game, such that $G$ aims to maximize the probability of $D$ making a mistake. In effect, $D$ acts as a critic, ensuring $G$ produces realistic reconstructions. The training process is as follows: i) Given a random normal 2D brain MRI slice, $x$, a square mask of length $\gamma$ is erased at a random location with the mask function $\Psi(x,\gamma)$. ii) The result of $\Psi(x,\gamma)$ is then passed to G for reconstruction, $G: \Psi(x,\gamma) \rightarrow \hat{x}$. Where $x, \Psi(x,\gamma), \hat{x} \in \mathbb{R}^{w \times h \times c}$. iii) $G$'s weights are then updated from a combined reconstruction and adversarial loss. To learn the contextual information of the normal training data, L1 loss is applied to $x$ and $G$'s reconstruction.
\begin{equation}
\mathcal{L}_{\text{rec}}=\underset{x \sim p_{x}}{\mathbb{E}}\left[||x - G(\Psi(x,\gamma))||_{1}\right].
\end{equation}
For $G$ to produce realistic reconstructions the following adversarial loss is incorporated to the overall learning objective:
\begin{equation}
\mathcal{L}_{\text{adv}}=\underset{x \sim p_{x}}{\mathbb{E}}[\log (1-D(G(\Psi(x,\gamma))))].
\end{equation}
Thus giving the complete learning objective for $G$:
\begin{equation}
\mathcal{L}_{\text{G}}=\lambda_{\text{rec}} \mathcal{L}_{\text{rec}} + \lambda_{\text{adv}} \mathcal{L}_{\text{adv}},
\end{equation}
where $\lambda_{\text{rec}}$ and $\lambda_{\text{adv}}$ are weighting parameters denoting the importance of reconstruction and adversarial loss respectively. iv) Finally, $D$'s weights are updated with the following:
\begin{equation}
\mathcal{L}_{\text{D}}=\underset{x \sim p_{x}}{\mathbb{E}}[\log D(x)]+\underset{x \sim p_{x}}{\mathbb{E}}[\log (1-D(G(\Psi(x,\gamma))))].
\end{equation}
To update the network parameters via the Adam optimizer algorithm, the \say{update} function is employed.
\begin{algorithm}
\label{alg:train}
\DontPrintSemicolon
\SetCommentSty{text}
\SetAlgoLined
 $\boldsymbol{\theta}_{\text{G}},\boldsymbol{\theta}_{\text{D}} =$ initialise network parameters\;
 $\lambda_{\text{rec}},\lambda_{\text{adv}} =$ initialise weighting parameters\;
 $\alpha_{\text{G}}, \alpha_{\text{D}}$ = initialise network learning rates\;
 $\gamma =$ initialise window size\;
 
 \Repeat{converged}{
    $\boldsymbol{\mathbf{x}} = $ random mini-batch from dataset\;
    $\hat{\mathbf{x}} = G(\Psi(\boldsymbol{\mathbf{x}},\gamma))$\;
    $\mathcal{L}_{\text{rec}} = ||\boldsymbol{\mathbf{x}} - \hat{\mathbf{x}}||_{1}$\;
    $\mathcal{L}_{\text{adv}} = \log(1 - D(\hat{\mathbf{x}}))$\;
    $\boldsymbol{\theta}_{\text{G}} = \small{\text{update}}(-\nabla_{\boldsymbol{\theta}_{\text{G}}} \lambda_{\text{rec}}\mathcal{L}_{\text{rec}} + \lambda_{\text{adv}}\mathcal{L}_{\text{adv}}, \alpha_{\text{G}})$\;
    $\boldsymbol{\theta}_{\text{D}}=\small{\text{update}}(\nabla_{\boldsymbol{\theta}_{\text{D}}}\log(D(\boldsymbol{\mathbf{x}})) + \log(1 - D(\hat{\mathbf{x}})), \alpha_{\text{D}})$\;
 }
 \caption{Train Inpainting Network}
\end{algorithm}

\emph{Architecture Details.} \hspace{9px}Inspired by UNet \cite{Ronneberger2015UNetCN}, $G$ inherits skip-connections such that each downsampling layer in the contracting path is concatenated to its corresponding up-sampling layer in the expansive path. These skip-connections allow for better quality reconstructions, increasing convergence speed and stabilizing training. In high-dimensional spaces, the density ratio estimation by the discriminator is often inaccurate and unstable during training, causing the generator to fail in learning the multimodal structure of the target distribution. Thus, for the discriminator we apply Spectral Normalization \cite{Miyato2018SpectralNF}, rescaling the weights with spectral norm $\sigma$ by the power iteration method.

\emph{Localization System.} \hspace{4px}To obtain predictions for all regions in $x$, the localization system employs the sliding window algorithm $\phi(x, \gamma, k)$, with a window size (square mask) of length $\gamma$ and a window step size of $k$ pixels. Each frame produced in $\phi(x, \gamma, k)$ is passed to the inpainting network for reconstruction. To reduce the impact of image contrast on reconstruction loss \cite{Berg2019UnsupervisedLO}, we apply minmax normalization $w(x)$ on $x$ and $G(\phi(x, \gamma, k))$, such that $w(x) : [\min(x), \max(x)]^{W \times H \times D} \mapsto [0, 255]^{W \times H \times D}$, where $\min(x)$ and $\max(x)$ are the minimum and maximum pixel values of $x$ respectively. After the normalization process, we compute the reconstruction (L1) loss between the normalized original window and normalized reconstructed window only. The result is a heatmap indicating regions that exhibit the most reconstruction loss.

To automate the segmentation process, we apply superpixel segmentation on the query slice $x$ and select the segment with the most reconstruction loss, based on the constructed heatmap. Superpixel based techniques segment an image into a collection of connected pixels which are similar in colour and texture. This allows us to capture common features in our slice and locate features our inpainting network failed to reconstruct. Although there exists many algorithms that perform superpixel segmentation, we choose to implement Felzenszwalb's efficient graph-based segmentation \cite{Felzenszwalb2004EfficientGI} as it segments tumours more accurately than other methods.

\emph{Dataset.} \hspace{4px}The Neurofeedback Skull-stripped (NFBS) \cite{Puccio2016ThePC} repository provides 125 manually skull-stripped T1-weighted MRI scans of normal brain tissue only, simplifying the inpainting task. Due to tumours being easier to identify, 2D input slices of size $256 \times 192$ are of the axial view of the brain. For evaluation, we utilise a tumorous dataset of 22 T1-weighted MRI scans provided by the Centre for Clinical Brain Sciences from the University of Edinburgh \cite{Pernet2016ANeuro}. The dataset consists of non-skull-stripped scans of 2D axial dimensions $256 \times 156$.

\emph{Preprocessing.} \hspace{6px}In addition to standardization and normalization for the tumour dataset, we utilise BrainSuite \cite{Shattuck2002BrainSuiteAA, Sandor1997SurfacebasedLO, Shattuck2001MagneticRI} for skull-stripping and bias field correction to rectify image intensity non-uniformities that are a result of magnetic field variations rather than anatomical differences. Furthermore, we add padding (black pixels) to each slice in the tumour dataset so that it matches the same dimensions as the normal dataset. Note, in every scan in the normal training set, the brain is already surrounded by black pixels, so not foreign to the inpainting network.

\section{Results}
\label{sec:results}

Evaluation was performed on $\gamma$ values of 8, 16, 32 and 64, with a $k$ step value of 4 pixels on separately trained networks. To evaluate the inpainting performance and segmentation accuracy, we extracted 100 slices from a normal test set \cite{Puccio2016ThePC} and 100 slices from the tumour dataset \cite{Pernet2016ANeuro} (where tumour is visible), respectively. The inpainting network was trained for 30,000 iterations in minibatches of size 16; learning rates for $G$ and $D$ were set to 0.0001 and 0.001; $\mathcal{L}_{\text{rec}}$ and $\mathcal{L}_{\text{adv}}$ were set to 50 and 1 respectively, as these settings empirically showed consistent optimal performance. We compared the segmentation performance with AnoGAN \cite{Schlegl2017UnsupervisedAD}, where AnoGAN \cite{Schlegl2017UnsupervisedAD} was trained for 60,000 iterations in minibatches of size 8.

\begin{figure}[hbt!]
    \centering
    \includegraphics[width=\linewidth]{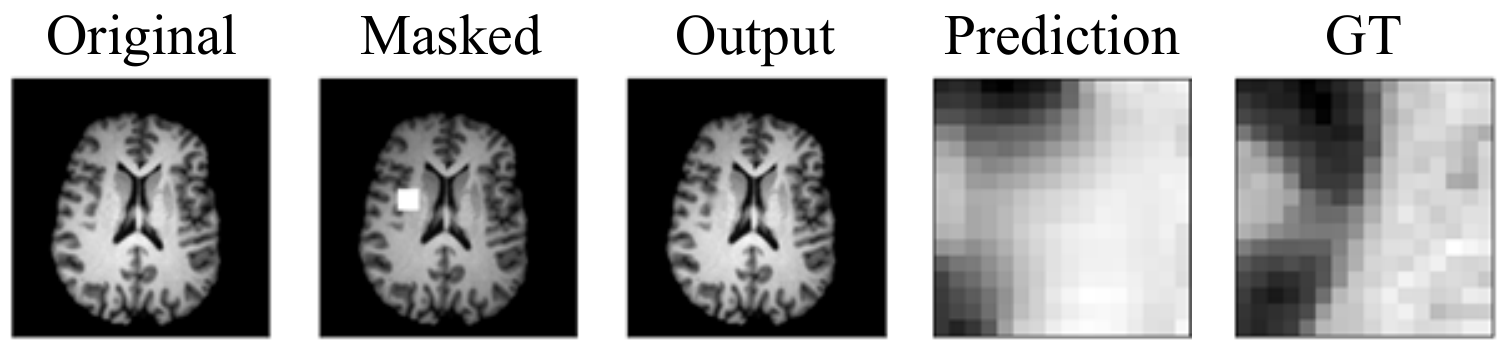}
    
    \vspace{0.25cm}
    
    \includegraphics[width=\linewidth]{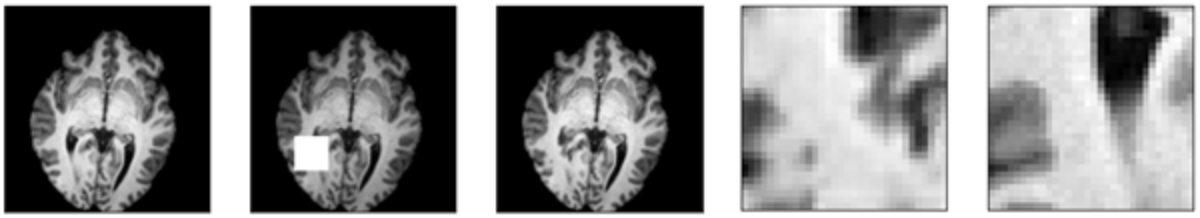}
    
    \vspace{0.25cm}
    
    \includegraphics[width=\linewidth]{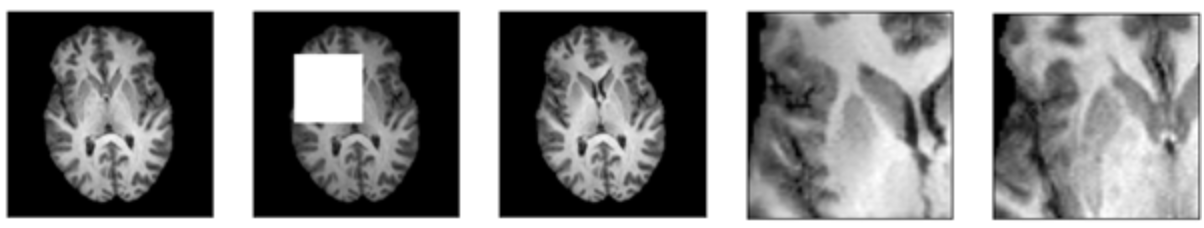}
    \caption{Inpainting results for window sizes 16, 32 and 64 respectively.}
    \label{fig:inpainting_different_size}
\end{figure}

As shown in figure \ref{fig:inpainting_different_size}, one can see how the problem definition changed according to the amount of data masked. For smaller window sizes, the inpainting network is only required to reconstruct a handful of pixels. However, with larger masks the network is required to reconstruct texture and structure. Additionally, due to adversarial training, the network produces reconstructions without blur.

\begin{table}[hbt!]
\centering
    \begin{adjustbox}{max width=0.49\textwidth}
        \begin{tabular}{cccc@{\qquad}}
            \toprule \toprule
            \text{} & PSNR ($\mu \pm \sigma$) & SSIM ($\mu \pm \sigma$) & Dice ($\mu \pm \sigma$)\\ \toprule\toprule 
            \makecell{Ours [8]} & 38.61 $\pm$ 3.159  & 0.99 $\pm$ 0.006 & 0.49 $\pm$ 0.396\\
            \makecell{Ours [16]} & 37.94 $\pm$ 4.664  & 0.98 $\pm$ 0.022 & 0.66 $\pm$ 0.346\\
            \makecell{Ours [32]} & 35.63 $\pm$ 5.413  & 0.97 $\pm$ 0.002 & \textbf{0.77} $\pm$ \textbf{0.176}\\
            \makecell{Ours [64]} & 31.66 $\pm$ 6.491  & 0.96 $\pm$ 0.003 & 0.55 $\pm$ 0.380\\
            \makecell{AnoGAN \cite{Schlegl2017UnsupervisedAD}} & -  & - & 0.38 $\pm$ 0.238\\
            \midrule\midrule
        \end{tabular}
    \end{adjustbox}
  \caption{Inpainting performace on unseen healthy samples and Dice scores.}
  \label{fig:inpainting_DICE_table}
\end{table}
\begin{figure}[hbt!]
    \centering
    \includegraphics[width=\linewidth]{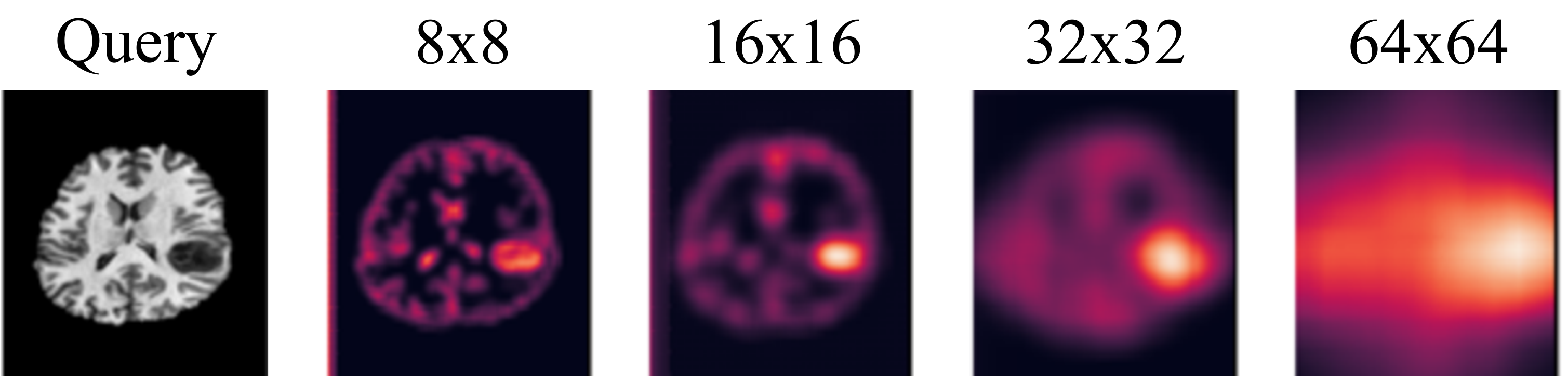}
    
    \vspace{0.16cm}
    
    \includegraphics[width=\linewidth]{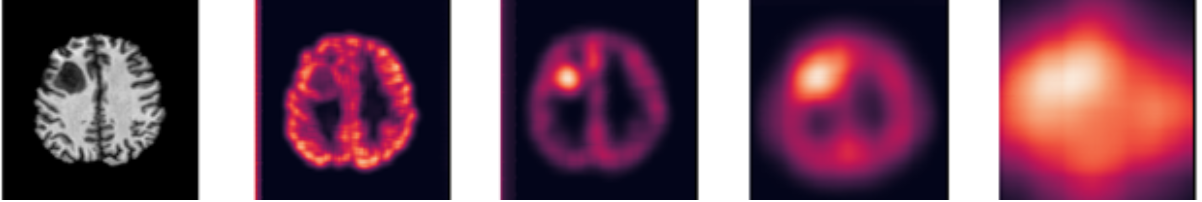}
    \caption{Heatmap for various window sizes.}
    \label{fig:heatmap}
\end{figure}

In figure \ref{fig:heatmap}, the heatmaps indicate higher reconstruction loss values where the tumour is present. Heatmaps produced from smaller window sizes performed better on healthy regions, whereas larger window sizes appear to localize the tumour more effectively. It is important that the window covers most if not all the tumour to distinguish the reconstruction performance with other healthy areas of the brain. However, if the size of the tumour is known beforehand, then choosing the \say{best-fit} window would be preferable.

\begin{figure}[hbt!]
    \vspace{0.5cm}
    
    \centering
    \begin{overpic}[width=0.485\textwidth]{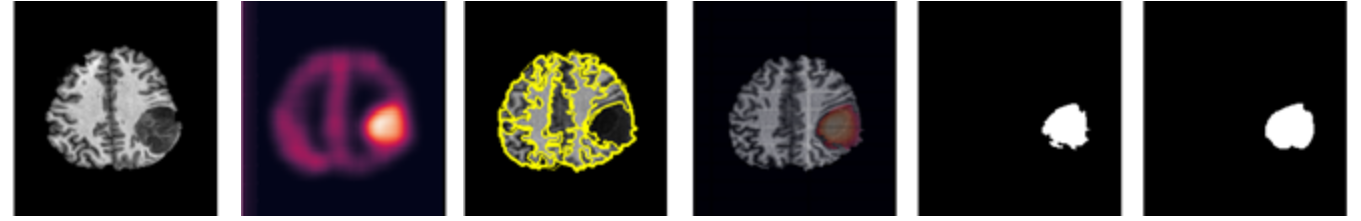}
    \put(6, 18.5){(a)}
    \put(23, 18.5){(b)}
    \put(39.5, 18.5){(c)}
    \put(56, 18.5){(d)}
    \put(73, 18.5){(e)}
    \put(89.5, 18.5){(f)}
    \end{overpic}

    \vspace{0.15cm}
    \includegraphics[width=0.99\linewidth]{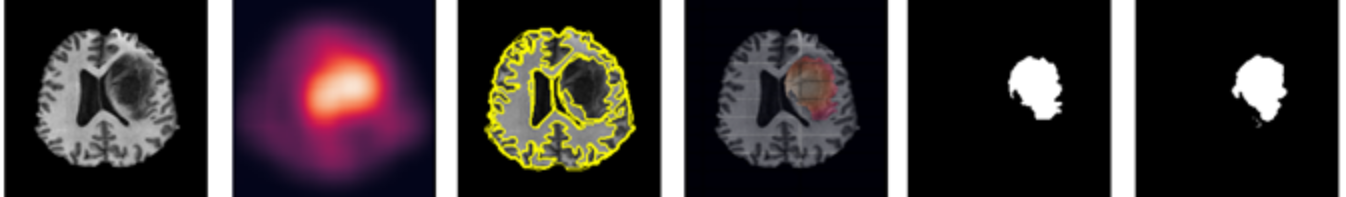}
    
    \vspace{0.15cm}
    \includegraphics[width=0.99\linewidth]{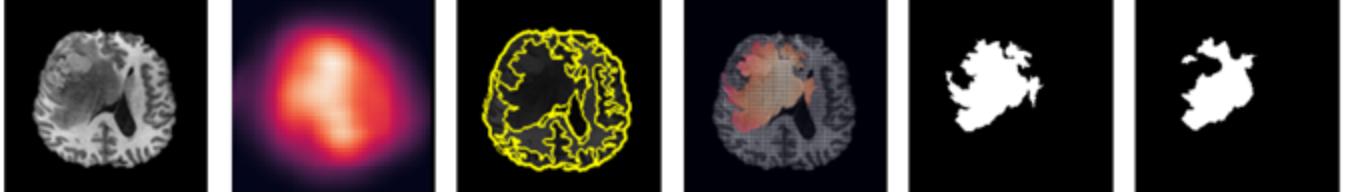}
    
    \caption{Segmentation Results - (a) Query, (b) Heatmap, (c) Superpixel, (d) Overlay, (e) Prediction, (f) Ground Truth}
    \label{fig:segmentation}
\end{figure}
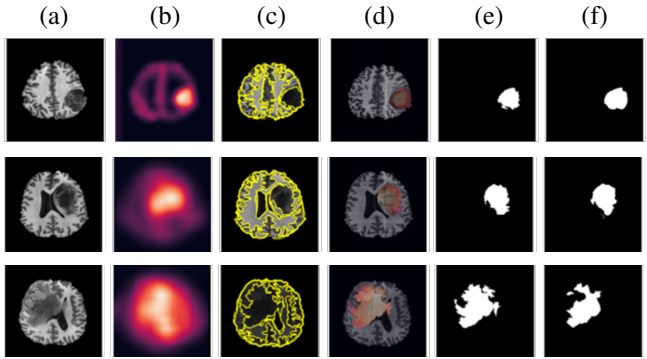
\emph{Segmentation Quality.} \hspace{0.3em} As shown in table \ref{fig:inpainting_DICE_table}, the proposed system outperforms AnoGAN \cite{Schlegl2017UnsupervisedAD}. A window of size 32 indicates the highest accuracy and consistency in handling various sized tumours, with a mean Dice score of 0.771 and of standard deviation 0.176. Other window sizes express high variability, performing well either on small or large sized tumours. The effect of using superpixel results in higher definition segmentations, allowing the capture of regular and more abstract tumours, which would not be possible with manual thresholding. Additionally, for evaluation, we trained a single network to reconstruct all values of  $\gamma$ (8, 16, 32 and 64). We obtained good reconstruction results, but this resulted in partial reconstructions of tumorous regions during the sliding window process, indicating a single optimal network for all tumour sizes is not possible with our method.

\section{Discussion}
\label{sec:discussion}
\begin{figure}[hbt!]
    \centering
    \includegraphics[width=\linewidth]{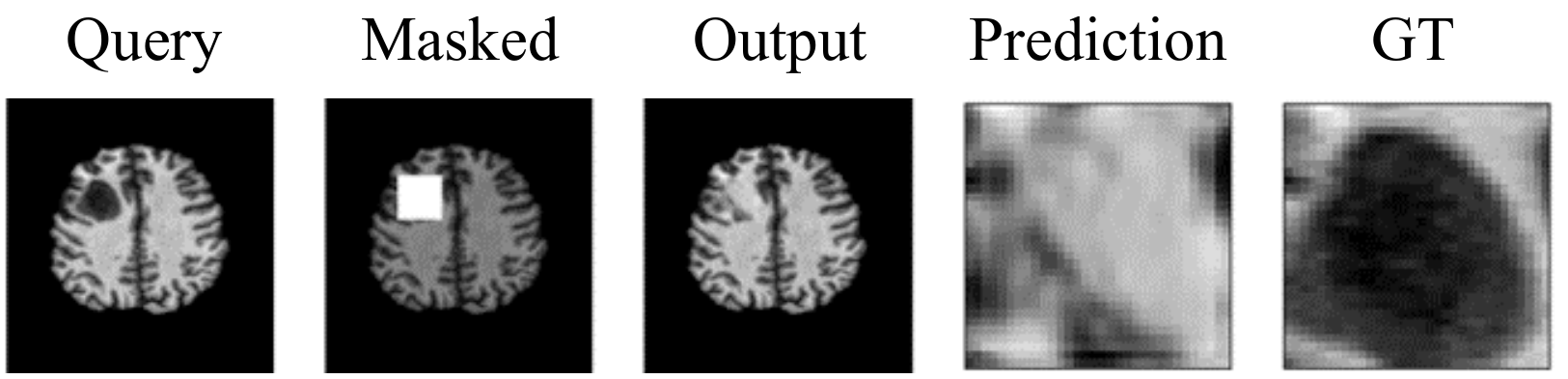}
    
    \vspace{0.25cm}
    
    \includegraphics[width=\linewidth]{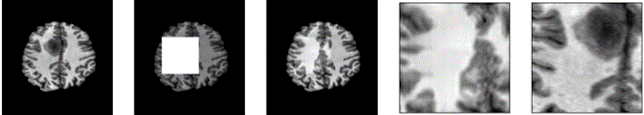}
    
    \vspace{0.25cm}
    
    \includegraphics[width=\linewidth]{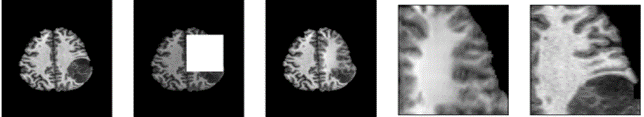}
    
    \caption{Examples of full and partial tumour coverage.}
    \label{fig:full_partial}
\end{figure}

Figure \ref{fig:full_partial} illustrates snapshots of the sliding window process. The inpainting network restores the masked region to what a normal brain would look like in that area. The inpainting network is naive and does not reconstruct the tumour when partially covered, due to computing L1 loss globally rather than just on the local masked region. 

Localization is only possible if the inpainting network performs well on non-anomalous regions, but fails otherwise. Due to high biological variability, non-anomalous regions may be viewed as anomalous. It is, therefore, important that the network is trained on a large and varied dataset. Additionally, the overall quality of segmentations depends on the superpixel algorithm. If tumours are not substantially different in colour and texture, no segmentation would be possible. 

\begin{figure}[hbt!]
  \centering
  
  \includegraphics[width=\linewidth]{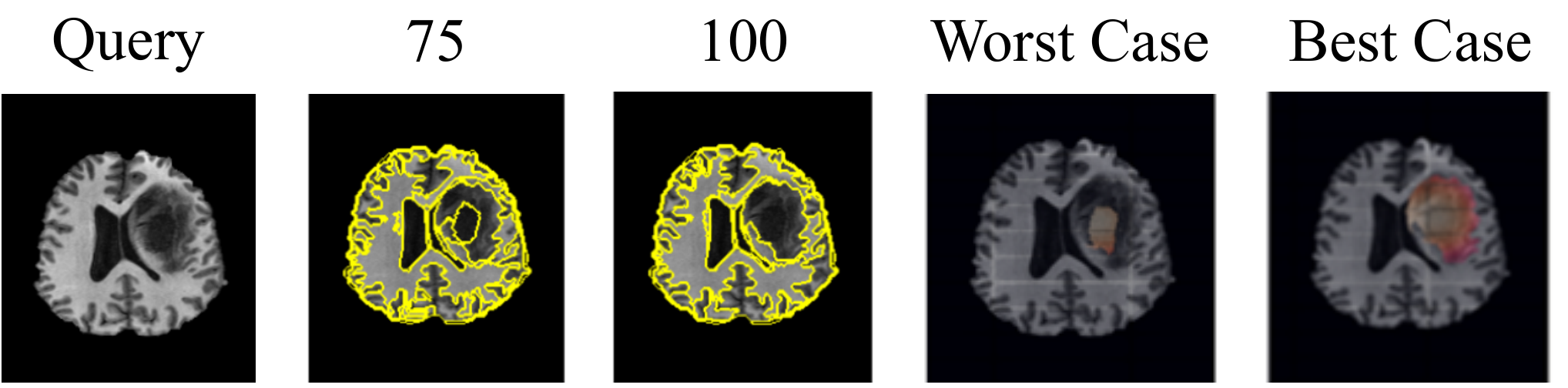}
    \caption{Superpixel segmentation with different scale values and possible outcomes.}
    \label{fig:k_value}
\end{figure}
In Felzenszwalb's algorithm \cite{Felzenszwalb2004EfficientGI}, the scale parameter influences the segment size. Although a constant scale value of 75 was used in the evaluation process, there were times when a larger or smaller value would have produced better segmentation results. Figure \ref{fig:k_value} illustrates how a scale value below 100 would result in the inner segment being selected. To prevent this, an appropriate scale value must be manually chosen before the segmentation process.

\section{Conclusion}
\label{sec:conclusion}

The proposed system for unsupervised anomaly detection in the application of T1-weighted brain MRI outperforms AnoGAN \cite{Schlegl2017UnsupervisedAD} for both the mean and standard deviation Dice scores, where the system accurately segments regular and abstract tumours of varying size. The main limitations are where normal segments surrounding the tumour are detected as false positives. However, the approach of an initial coarse unsupervised region-based segmentation strategy, followed by a separate refinement stage, has proved to be effective in the application of unsupervised medical segmentation.

\section{Compliance with Ethical Standards}
\label{sec:ethics}

The training data used \cite{Puccio2016ThePC} in this research study was conducted retrospectively using human subject data made open access by the Nathan Kline Institute. The data was acquired in compliance with the Health Insurance Portability Act (HIPPA). The evaluation data used \cite{Pernet2016ANeuro} was also made open access by the Centre for Clinical Brain Sciences from the University of Edinburgh.

\bibliographystyle{IEEEbib}
\bibliography{root.bbl}
\end{document}